\documentclass[twocolumn,aps,prl,showpacs,amsmath,amssymb]{revtex4-1}
\usepackage{epsfig}
\usepackage{graphicx}
\usepackage{dcolumn}
\usepackage{bm}
\usepackage[colorlinks=true,dvipdfm]{hyperref}

\begin{document}
\title{Scaling behavior of quantum critical relaxation dynamics in a heat bath}
\author{Shuai Yin}
\author{Chung-Yu Lo}
\author{Pochung Chen}
\affiliation{Department of physics, National Tsing Hua university, Hsinchu 30013, Taiwan}
\date{\today}

\begin{abstract}

We study the scaling behavior of the relaxation dynamics to thermal equilibrium when a quantum system is near the quantum critical point. In particular, we investigate systems whose relaxation dynamics is described by a Lindblad master equation. We find that the universal scaling behavior not only exhibits in the equilibrium stage at the long-time limit, but also manifests itself in the non-equilibrium relaxation process. While the critical behavior is dictated by the low-lying energy levels of the Hamiltonian, the dissipative part in the Lindblad equation also plays important roles in two aspects: First, the dissipative part makes the high energy levels decay fast after which the universal behavior controlled by the low-lying modes emerges. Second, the dissipation rate gives rise to a time scale that affects the scaling behavior. We confirm our theory by solving the Lindblad equation for the one-dimensional transverse-field Ising model.
\end{abstract}

\maketitle
Quantum critical dynamics near the quantum phase transitions has attracted a lot of attentions recently~\cite{Dz,Pol}. These studies are motivated by the progress of detecting and manipulating dynamic phenomena in cold atom systems~\cite{Greiner,Meinert}. While it is well known that a $D$-dimensional quantum phase transition can be mapped into a corresponding $D+1$-dimensional classical phase transition~\cite{Sachdev}, to what extent the non-equilibrium scaling behavior is shared by the classical and quantum phase transitions, however, is still an open question. One positive example is the Kibble-Zurek mechanism, which was originally proposed in the context of classical phase transitions to describe the scaling of the production of the topological defects in a driven process crossing the critical point~\cite{Zurek,Kibble}. Recently it has been proved that the same mechanism is also applicable in quantum critical dynamics~\cite{Dz,Pol}. On the other hand, the relaxation dynamics after a sudden quench is an example, in which the critical behaviors differ in classical and quantum cases. In classical phase transitions, the dissipative nature of thermodynamics makes the high-energy modes decay fast~\cite{Hohen}. Consequently after a transient time, the critical system enters the universal stage dominated by low-energy modes. In this universal stage, observable demonstrates power-law decay behavior. This is the well-known critical-slowing-down behavior~\cite{Cala}. In contrast for quantum dynamics of closed systems, the distribution in different energy levels is completely determined by the initial state and is unchanged during the evolution due to the unitary property of the Schr\"{o}dinger equation. As a result, the relaxation dynamics does not show critical-slowing-down in closed quantum systems~\cite{Dz,Pol,Barm}.

Although recent experimental technologies can isolate a quantum system quite well from the environment~\cite{Newton}, in general, quantum systems are inevitably coupled to the environment in real situations. This gives rise to the concept of open quantum systems~\cite{Breuer,Weiss}. The most common case is a quantum system, which is coupled to an infinitely large heat bath at temperature $T$. According to the zeroth law of thermodynamics, the system will reach a thermal equilibrium state characterized by the same temperature $T$. While for close system it is known that the equilibrium state dictates the equilibrium scaling behavior when the parameters of the system Hamiltonian are near their critical values, the scaling properties of the open quantum critical system's relaxation process approaching the thermal equilibrium state remain elusive. This work is also motivated by another important issue in quantum phase transition, that is, the quantum critical dynamics at finite temperatures~\cite{Sachdev,Sachdev3,Sachdev2}. Due to the interplay between the thermal fluctuation and the long-range entanglement this issue is difficult to tackle. Effective methods, both in analytic and numerical aspects, are hence desired to understand the dynamic quantum criticality at finite temperatures. For closed systems, both relaxation dynamics and driven quantum critical dynamics have been investigated~\cite{Sachdev,Polk,Soti,Grit,Deng}. For open systems connected to a heat bath, it has been shown that both the temperature~\cite{Patane1,Patane2,Yin} and the dissipation rate~\cite{Yin} affect the scaling behavior in driven dynamics. Relaxation dynamics has also been studied for the case of thermal quench in the anisotropic Ising model interacting with a bosonic bath~\cite{Patane3}. However, a more comprehensive understanding is still called for.

To shed some light on these issues in this work we address the following questions: Does the relaxation dynamics in the presence of a heat bath demonstrate scaling behavior near the quantum critical point? Furthermore, how to characterize it if such a scaling behavior does exist? To answer these questions, we study the relaxation dynamics described by the master equation of the Lindblad form~\cite{Attal,Breuer,Weiss}. We shall show that the non-equilibrium relaxation process demonstrates scaling behaviors. The reason for the emergence of these non-equilibrium scaling properties is closely related to the dissipative nature of the dynamics. Because of the dissipation, the high-energy modes decay fast and the surviving low-energy modes give rise to the universal scaling behavior. Based on the scale transformation in the Lindblad equation, we propose a scaling theory to describe the universal behavior in the relaxation dynamics. We verify the scaling theory by solving the Lindblad equation for the one-dimensional ($1$D) transverse-field Ising model. We will also discuss the condition for the emergence of the critical-slowing-down induced by the dissipation.


The master equation in the Lindblad form describes the dynamics of an open quantum system which is weakly coupled to an environment~\cite{Breuer,Weiss}. For different environments, the master equation could have different operators in it~\cite{Orszag,Znidari,Fubini}. When the environment is a heat bath, the Lindblad equation is~\cite{Attal,Breuer,Weiss}
\begin{eqnarray}
\begin{split}
\frac{\partial \rho}{\partial
t}=&-i[\mathcal H,\rho]-\sum_{m\neq l}c_{lm}W_{l\rightarrow m}(V_{l\rightarrow m}^\dagger V_{l\rightarrow m} \rho\\&+\rho V_{l\rightarrow m}^\dagger V_{l\rightarrow m}-2V_{l\rightarrow m} \rho V_{l\rightarrow m}^\dagger), \label{Lind}
\end{split}
\end{eqnarray}
in which $\rho$ is the density matrix of the system, $\mathcal H$ is the Hamiltonian of the system, $c_{lm}\equiv c_{ml}$ is the dissipation rate, which generally depends on the energy levels, $V_{l\rightarrow m}\equiv|m\rangle \langle l|$ is the jump matrix from the $m$th energy level to the $l$th one, and $W_{l\rightarrow m}$ is the transition probability satisfying the detailed balance condition, $W_{l\rightarrow m}/W_{m\rightarrow l}={\rm exp}[-(E_m-E_l)/T]$. This condition ensures that thermal state will be reached in the long time limit, independent of the detailed form of $W_{l\rightarrow m}$.

In addition, the Lindblad equation has an adjoint form. For any local operator $Y$ whose time-dependent expectation value is $\langle Y\rangle(t)\equiv{\rm Tr}[\rho(t)Y]$, it is convenient to define a time-dependent operator $Y(t)$ which satisfies ${\rm Tr}[\rho(t)Y]\equiv {\rm Tr}[\rho(0)Y(t)]$. It has been proved that $Y(t)$ satisfies the adjoint Lindblad equation ~\cite{Breuer}
\begin{eqnarray}
\begin{split}
\frac{\partial Y(t)}{\partial t}=&-i[{\mathcal H},Y(t)]-\sum_{m\neq l}c_{lm}W_{l\rightarrow m}[V_{l\rightarrow m}^\dagger V_{l\rightarrow m} Y(t)\\&+Y(t) V_{l\rightarrow m}^\dagger V_{l\rightarrow m}-2V_{l\rightarrow m}^\dagger Y(t)V_{l\rightarrow m}]. \label{Lindadj}
\end{split}
\end{eqnarray}

In the long-time limit, i.e., $t\rightarrow \infty$, the steady solution of Eq.~(\ref{Lind}) is the canonical distribution, $\rho_E={\rm exp}(-{\mathcal H}/T)/{\rm Tr}[{\rm exp}(-{\mathcal H}/T)]$~\cite{Attal,Breuer,Weiss}. For Eq.~(\ref{Lindadj}), the equilibrium solution gives the expectation value of $Y$ in the canonical distribution. That is $\langle Y\rangle_E={\rm Tr}(\rho_E Y)$~\cite{Breuer}. We note that these equilibrium solutions are independent of the initial conditions and the dissipation rate. Therefore, the equilibrium scaling behavior, controlled by this distribution, is also independent of the initial condition and the dissipation rate.

The dynamics described by the Eqs.~(\ref{Lind}) and (\ref{Lindadj}) includes contributions from both the quantum and classical thermal fluctuations. The first part in the right hand side of Eq.~(\ref{Lind}) shows the quantum unitary evolution; while the second part of the right hand side gives the master equation describing the classical stochastic process~\cite{Cardy}. In particular, when we set the temperature of the environment to be zero, Eqs.~(\ref{Lind}) and (\ref{Lindadj}) describe the dynamics of the spontaneous emission process.



To study the non-equilibrium scaling behavior described by Eq.~(\ref{Lind}), we impose a scale transformation with a scaling factor $b$ on Eq.~(\ref{Lind}). Under this transformation, $t\rightarrow \tilde{t}=tb^{-z}$, $\mathcal H\rightarrow \tilde{\mathcal H}=\mathcal Hb^z$, $T\rightarrow \tilde{T}=Tb^z$, where $z$ is the dynamic exponent, while the dimensionless quantities, $\rho$, $V_{l\rightarrow m}$ and $W_{l\rightarrow m}$ are unchanged. Consequently, Eq.~(\ref{Lind}) becomes
\begin{eqnarray}
\begin{split}
\frac{\partial \rho}{\partial
\tilde{t}}=&-i[\tilde{\mathcal H},\rho]-b^z\sum_{m\neq l}c_{lm}W_{l\rightarrow m}(V_{l\rightarrow m}^\dagger V_{l\rightarrow m} \rho\\&+\rho V_{l\rightarrow m}^\dagger V_{l\rightarrow m}-2V_{l\rightarrow m} \rho V_{l\rightarrow m}^\dagger). \label{Lind2}
\end{split}
\end{eqnarray}
The scale invariance of Eq.~(\ref{Lind}) requires the scale transformation of the dissipation rate to be $\tilde{c}_{lm}\equiv c_{lm}b^z$. This demonstrates that $c_{lm}$ has the dimension of $z$. As the dissipation rate does not affect the equilibrium scaling behavior as we discussed above, it is called a \textit{dynamically relevant} scaling variable. This property of the dissipation rate has been shown in the driven critical dynamics~\cite{Yin}.

To explicitly show the scaling behavior of the relaxation dynamics, we solve Eq.~(\ref{Lind}) in the low-lying energy levels at zero temperature. In this case, Eq.~(\ref{Lind}) actually characterizes the dynamics of the spontaneous emission. In general due to the dissipative nature of Eq.~(\ref{Lind}), after a transient time scale the high-energy modes decay fast and only the low-lying energy modes survive. Under this condition, the diagonal part of Eq.~(\ref{Lind}) is
\begin{equation}
\frac{\partial \rho_{ll}}{\partial t}=-c_{l0}\rho_{ll}, \label{diagLind}
\end{equation}
In this part, we find that the unitary commutator in Eq.~(\ref{Lind}) has no contribution. The off-diagonal part shows
\begin{equation}
\frac{\partial \rho_{l0}}{\partial t}=-i(E_l-E_0) \rho_{l0}-\frac{1}{2}c_{l0}\rho_{l0}, \label{offdiagLind}
\end{equation}
and the solution is
\begin{equation}
\rho_{l0}(t)=\rho_{l0}(0){\rm exp}[-i(E_l-E_0) t-\frac{1}{2}c_{l0}t]. \label{offdiagLind1}
\end{equation}
Equations~(\ref{diagLind}), (\ref{offdiagLind}) and (\ref{offdiagLind1}) explicitly demonstrate that at the zero temperature, non-equilibrium universal behavior of the spontaneous emission is dominated by the interplay of the following two time scales: One is the inverse of the energy gap, $(E_l-E_0)^{-1}$, the other is the dissipation rate $c_{l0}^{-1}$. Equation~(\ref{offdiagLind1}) also shows that the off-diagonal elements of the density matrix demonstrate damped oscillation behavior. The period of oscillation is the inverse of energy gap, while the damped time scale is the inverse of the dissipation rate.

Near the critical point, for an operator $Y$, whose transformation is $Y\rightarrow \tilde{Y}=Yb^s$ under a scale transformation with a scaling factor $b$, $\tilde{Y}(t)$ satisfies the rescaled adjoint Lindblad equation
\begin{eqnarray}
\begin{split}
\frac{\partial \tilde{Y}(t)}{\partial
\tilde{t}}=&-i[\tilde{\mathcal H},\tilde{Y}(t)]-\sum_{m\neq l}\tilde{c}_{lm}W_{l\rightarrow m}[V_{l\rightarrow m}^\dagger V_{l\rightarrow m} \tilde{Y}(t)\\&+\tilde{Y}(t) V_{l\rightarrow m}^\dagger V_{l\rightarrow m}-2V_{l\rightarrow m}^\dagger \tilde{Y}(t)V_{l\rightarrow m}]. \label{Lind3}
\end{split}
\end{eqnarray}
Accordingly, we obtain scale transformation of $\langle Y\rangle$,
\begin{eqnarray}
\begin{split}
\langle Y\rangle(t,{\mathcal H},T,c_{lm})=b^{-s}\langle \tilde{Y}\rangle(tb^{-z},{\mathcal H}b^z,Tb^z,c_{lm}b^z). \label{scatra}
\end{split}
\end{eqnarray}
We note that in Eq.~(\ref{scatra}), $c_{lm}$ only involves the low energy contributions. Equation~(\ref{scatra}) describes the scaling behavior of $\langle Y\rangle$ in the universal relaxation stage. Comparing with the classical critical dynamics, which generally only includes time $t$ as its additional scaling variable, here we find in the present class of relaxation quantum critical dynamics, the dissipation rate must be a scaling variable. After entering the equilibrium stage, $c_{lm}$ must vanish as the equilibrium final state is independence of the dissipation rate and Eq.~(\ref{scatra}) recovers the equilibrium scale transformation.

In particular, we consider a Hamiltonian with $g$ being the distance to the critical point, $h$ being the symmetry breaking field, and $L$ being the lattice size. For the order parameter $M$, when the evolution starts with a saturated state, the order parameter should satisfies
\begin{eqnarray}
\begin{split}
M(t,g,&h,T,c_{lm},L)=\\&b^{-\beta/\nu}\tilde{M}(tb^{-z},gb^{1/\nu},hb^{\beta\delta/\nu},Tb^z,c_{lm}b^z,Lb^{-1}), \label{scaop}
\end{split}
\end{eqnarray}
in which $\beta$ is defined in $M\sim (-g)^{\beta}$ for $g<0$ and $h=0$~\cite{Sachdev},  $\delta$ is defined in $M\sim h^{1/\delta}$ at $g=0$~\cite{Sachdev}, and $\nu$ is defined in $\xi\sim |g|^{-\nu}$ ($\xi$ is the correlation length)~\cite{Sachdev}. In the following, we will consider the finite-size scaling (FSS)~\cite{Sondhi}. By assuming $Lb^{-1}=1$, we get the FSS form of the order parameter $M$,
\begin{eqnarray}
\begin{split}
M(t,g,&h,T,c_{lm},L)=\\&L^{-\beta/\nu}f(tL^{-z},gL^{1/\nu},hL^{\beta\delta/\nu},TL^z,c_{lm}L^z). \label{scaop1}
\end{split}
\end{eqnarray}

To verify the scaling theory, we take the $1$D transverse-field Ising model as an example. The Hamiltonian is \cite{Sachdev}
\begin{equation}
\mathcal{H}=-\sum_{i=1}^{L-1} \sigma_{i}^z\sigma_{i+1}^z-h_{\rm{x}}\sum_{i=1}^L\sigma_i^x-h\sum_{i=1}^L\sigma_i^z,
\label{modelI}
\end{equation}
where $\sigma_{i}^x$ and $\sigma_{i}^x$ are the Pauli matrixes at site $i$ in the $x$ and $z$ directions respectively, while $h_{\rm x}$ and $h$ are the transverse-field and symmetry-breaking field respectively. The critical point of model (\ref{modelI}) is located at $h_{\rm x}=h_{\rm xc}\equiv 1$~\cite{Sachdev}. The exact critical exponents are $\beta=1/8$, $\nu=1$, $\delta=15$, and $z=1$~\cite{Sachdev}. This model has been realized in CoNb$_2$O$_6$~\cite{Coldea}.

We solve directly Eq.~(\ref{Lind}) by using the finite difference method with periodic boundary condition. The time interval is chosen as $0.0004$. Smaller intervals have been checked to produce no appreciable changes. The lattice sizes used are $L=6$, $7$, $8$, $9$, and $10$. It has been shown that the critical dynamics for these sizes can be described by the finite-size scaling~\cite{Yin}. We shall choose $W_{l\rightarrow m}$ as $W_{l\rightarrow m}={\rm exp}[-(E_m-E_0)/T]$~\cite{Yin}. Since $W_{l\rightarrow m}$ is dimensionless, its detailed form will not affect the universal scaling properties. To eliminate initial effects induced by the initial magnetization, we also choose a  initial state with $M_0=1$ since the saturated magnetization is an apparent fixed point of $M_0$~\cite{Cala}. In the following we use two specific forms of the dissipation rate to verify the proposed scaling behaviors.

\begin{figure}
  \centerline{\epsfig{file=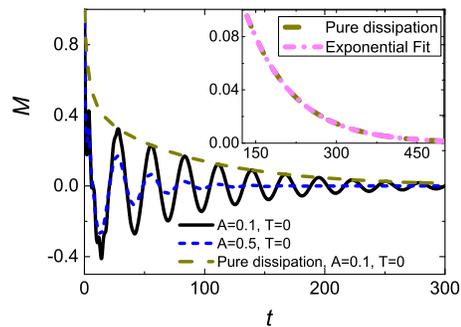,width=0.7\columnwidth}}
  \caption{\label{spt1} (Color online) Evolution of $M$ for $L=7$ in various cases. Initial magnetization is chosen as $M_0=1$. Inset: Fitting of curve for the pure dissipative case.}
\end{figure}

\textit{Case A: $c_{lm}$ depends on energy levels}. In this case we choose $c_{lm}=A|E_l-E_m|$. This is motivated by the fact that in general this term is the lowest order term for the dependence of the dissipation rate on the energy levels. Near the critical point, as the energy gap tends to zero, this linear term will dominate because the higher-order terms, if they exist, will be much smaller than this linear term. Furthermore, even if the leading term is a higher-order term, the scaling behavior will be similar to the present situation, and the only modification is to take the dimension of the coefficient into account.

Figure~\ref{spt1} shows the evolution of the order parameter for $L=7$ at $g=0$, $h=0$, and $T=0$. For a finite-size system, a energy gap will be induced by the lattice size. At zero temperature, the main low-lying modes are the degenerate first excited states and the ground state. Consequently, the scaling behavior is dominated by $\Delta\equiv E_1-E_0$ and $c_{10}$. From Fig.~\ref{spt1} we find that the period is indeed $2\pi/\Delta$. We also find that the amplitude of the curve decay according to the pure dissipative dynamics, which is obtained by ignoring the unitary part in the Lindblad equation. By fitting the pure dissipative curve, we find that the curve is proportional to ${\rm exp}(-c_{10}t/2)$ after $t\simeq 100$. These results confirm that the dynamics is described by Eq.~(\ref{offdiagLind1}). Additionally by comparing
the curves for  different coefficients $A$ in dissipation rates, we find that the period of the oscillation is independent of the dissipation rate. This is because the the oscillating effects is only determined by the unitary part of the Lindblad equation.

\begin{figure}
  \centerline{\epsfig{file=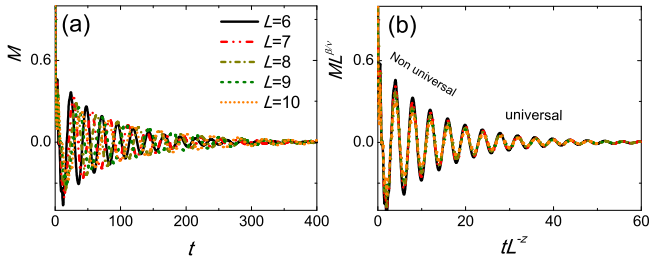,width=1.0\columnwidth}}
  \caption{\label{spt2} (Color online) At the zero temperature, the curves $M$ versus $t$ for various lattice size in (a) overlap well in (b) when $M$ and $t$ are rescaled with $L$. $A=0.1$ for different lattice sizes.}
\end{figure}

In figure~\ref{spt2}(a) we show the evolution of the order parameter of various lattice sizes at zero temperature, $h=0$, and $A=0.1$. We observe that the period of the oscillation increases as the lattice size increases. This is because the period is proportional to $1/\Delta\sim L^z$. In figure~\ref{spt2}(b) we plot the rescaled order parameter $M L^{\beta/\nu}$ as a function of the rescaled time $tL^z$. We find that there is a transient time within which the collapse is less than ideal. This is because the high-energy modes make significant contributions in this non-universal stage. After that an excellent collapse is observed and  the universal behavior, controlled by the low-energy modes, emerges. These results clearly confirm the scaling form of Eq.~(\ref{scaop1}) at $T=0$ and parameters are at their critical values.

\begin{figure}
  \centerline{\epsfig{file=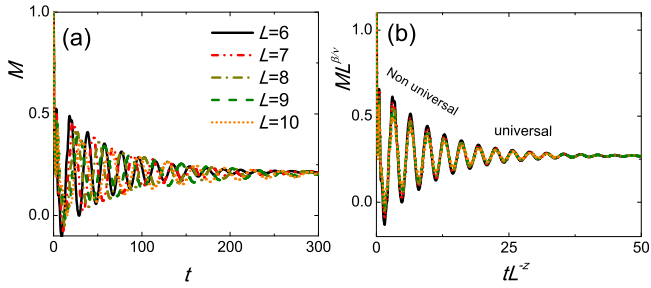,width=1.0\columnwidth}}
  \caption{\label{spt4} (Color online) In the presence of symmetry-breaking field, the curves $M$ versus $t$ for various lattice size and fixed $TL^z\simeq 2.1$ and $hL^{\beta\delta/\nu}\simeq 0.573$ in (a) overlap well in (b) when $M$ and $t$ are rescaled with $L$. $A=0.1$ for different lattice sizes.}
\end{figure}

We next investigate if Eq.~(\ref{scaop1}) is still robust when the temperature is nonzero and the parameters in the Hamiltonian deviate from their critical values. In Figure~\ref{spt4}(a) and Figure~\ref{spt4}(b) we plot the evolution of the order parameter without and with rescaling respectively for a fixed  $TL^z$ and $hL^{\beta\delta/\nu}$. Similar to the case of $T=0$, excellent collapse is reached in the universal regime after the transient time. Since when $h\neq 0$, the integrability of the transverse-field Ising model breaks down, our results also demonstrate the  integrability of the model is irrelevant to the present scaling theory.

\begin{figure}
  \centerline{\epsfig{file=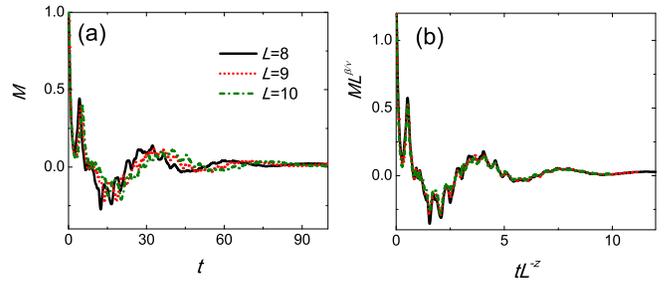,width=1.0\columnwidth}}
  \caption{\label{spt5} (Color online) (a) The curves $M$ versus $t$ for various lattice size and fixed $TL^z\simeq 2.1$, $hL^{\beta\delta/\nu}\simeq 0.573$ and $cL^z\simeq 1.2$. (b) After rescaling, the outlines of these curves collapse well, but the high frequency part cannot collapse.}
\end{figure}

\textit{Case B: $c_{lm}$ is a constant $c$}. When the dissipation rate is a constant, universal behavior controlled by the low energy levels is still described by Eq.~(\ref{scaop}). However, in this particular case, all excited modes decay at the same rate. Consequently when the system is near the equilibrium state, while the low energy levels contribute the scaling behavior, the high energy modes also take part in the dynamics. In this scenario we call the relaxation dynamics showing a {\it weak} scaling behavior. Figure~\ref{spt5} shows the relaxation dynamics of model~(\ref{modelI}) without and with rescaling for fixed $hL^{\beta\delta/\nu}$, $TL^z$ and $cL^z$. According to the discussion above, the oscillation of the outlines is dictated by the low-energy modes, which are controlled by the finite-size effects. On the other hand the high-frequency oscillations riding on the outline are brought by the high-energy levels. Although the outlines collapse together, the high-momentum oscillations do not match with each other after rescaling. This illustrates the weak scaling behavior.

Finally, we discuss the condition for the emergent critical-slowing-down of the relaxation process in the thermodynamic limit. When $L\rightarrow\infty$, $T=0$, $h=0$ and $g=0$, Eq.~(\ref{scaop}) becomes
\begin{eqnarray}
\begin{split}
M(t,c_{l0})=b^{-\beta/\nu}M(tb^{-z},c_{l0}b^z). \label{scaop2}
\end{split}
\end{eqnarray}
Although Eq.~(\ref{scaop2}) gives a general description of the critical dynamics in the thermodynamic limit at the critical point, the relaxation behavior will depend on the detailed information of $c_{l0}$. To see this, we return to Eqs.~(\ref{diagLind}) and (\ref{offdiagLind}) and consider a relaxation dynamics with $c_{l0}=A|E_l-E_0|$. For a specific low energy mode, $w$, its energy scale is $\Delta_w=E_w-E_0$. This mode decays in a time scale $t_w\sim \Delta_w^{-1}$ after which the system will enter the next stage dominated by the modes with lower energies. Since in the thermodynamics limit the gap between the first excited state and the ground state approaches zero at the critical point, the decay time for the first excited state will tend to infinity. We call this phenomena the \textit{critical-slowing-down induced by the dissipation}. This phenomena also manifests itself in the oscillation behavior, as the period of oscillation will become longer and longer since lower and lower energy modes dominate in the relaxation process. However, when the dissipation rate is a constant $c$, all high energy modes decay in the same time scale and the system reaches the ground mode in a scale characterized by $c$. Consequently there is no critical slowing down for the case of a constant dissipation rate.

In summary, we have studied the scaling behavior of the relaxation dynamics described by the Lindblad equation, at both zero and finite temperatures. By analysing the scale transformation in the Lindblad equation and its adjoint form, we have proposed a general scaling theory to describe the relaxation dynamics in the universal stage dominated by the low-lying energy levels. In this scaling theory, we have explored the role played by the dissipation rate. We verify the scaling theory by solving the Lindblad equation analytically and numerically. We have also argued that the critical-slowing-down induced by the dissipation can appear for a class form of the dissipation but not for the case of a constant dissipation rate.

We acknowledge the support by MOST of Taiwan through Grant No. 104-2628-M-007-005-MY3. We also acknowledge the support from National Center for Theoretical Science (NCTS) of Taiwan.


\begin{thebibliography}{99}
\bibitem{Dz} J. Dziarmaga, Adv. Phys. {\bf 59}, 1063 (2010).
\bibitem{Pol} A. Polkovnikov, K. Sengupta, A. Silva, and M. Vengalattore, Rev. Mod. Phys. {\bf 83}, 863 (2011).
\bibitem{Greiner} M. Greiner, O. Mandel, T. Esslinger, T. W. H\"{a}nsch, and I. Bloch, Nature (London) {\bf 415}, 39 (2002).
\bibitem{Meinert} F. Meinert, M. J. Mark, E. Kirilov, K. Lauber, P. Weinmann, A. J. Daley, and H.-C. N\"{a}gerl, Phys. Rev. Lett. {\bf 111}, 053003 (2013).
\bibitem{Sachdev} S. Sachdev, \textit{Quantum Phase Transitions}(Cambridge University Press, 1999).
\bibitem{Zurek} W. H. Zurek, Nature (London) {\bf 317}, 505 (1985).
\bibitem{Kibble} T. Kibble, J Phys. A {\bf 9}, 1387 (1976). 
\bibitem{Hohen}P. C. Hohenberg and B. I. Halperin, Rev. Mod. Phys. {\bf 49}, 435 (1977).
\bibitem{Cala}P. Calabrese and A. Gambassi, J. Stat. Mech.: Theory Exp. (2005) P04010.
\bibitem{Barm}P. Barmettler, M. Punk, V. Gritsev, E. Demler, and E. Altman, Phys. Rev. Lett. {\bf 102}, 130603 (2009).
\bibitem{Newton}T. Kinoshita, T. Wenger, and D. S. Weiss, Nature {\bf 440}, 900 (2006).
\bibitem{Breuer} H. P. Breuer and F. Petruccione, {\it The Theory of Open Quantum Systems} (Oxford University Press, Oxford, U.K., 2002).
\bibitem{Weiss}U. Weiss, {\it Quantum Dissipative Systems, 3rd ed.} (World Scientific, Singapore, 2008).
\bibitem{Sachdev3}S. Sachdev and A. P. Young, Phys. Rev. Lett. {\bf 78}, 2220 (1997).
\bibitem{Sachdev2}S. Sachdev, in \textit{Recent Progress in Many-Body Theories}, edited by R. F. Bishop et al. (World Scientific, Singapore, 2002).
\bibitem{Polk} A. Polkovnikov and V. Gritsev, Nat. Phys. {\bf 4}, 477 (2008).
\bibitem{Soti} S. Sotiriadis, P. Calabrese, and J. Cardy, Europhy. Lett. {\bf 87}, 20002 (2009).
\bibitem{Grit} V. Gritsev and A. Polkovnikov, in {\it Understanding Quantum Phase Transitions}, edited L. D. Carr (Taylor \& Francis, Boca Raton, FL, 2010).
\bibitem{Deng} S. Deng, G. Ortiz, and L. Viola, Phys. Rev. B 83, 094304 (2011).
\bibitem{Yin} S. Yin, P. Mai, and F. Zhong, Phys. Rev. B {\bf 89}, 094108 (2014).
\bibitem{Patane1}D. Patan\`{e}, A. Silva, L. Amico, R. Fazio, and G. E. Santoro, Phys. Rev. Lett. {\bf 101}, 175701 (2008).
\bibitem{Patane2}D. Patan\`{e}, L. Amico, A. Silva, R. Fazio, and G. E. Santoro, Phys. Rev. B {\bf 80}, 024302 (2009).
\bibitem{Patane3}D. Patan\`{e}, A. Silva, F. Sols, and L. Amico, Phys. Rev. Lett. {\bf 102}, 245701 (2009).
\bibitem{Attal}S. Attal and A. Joye, J. Funct. Anal. {\bf 247}, 253 (2007).
\bibitem{Orszag} M. Orszag, {\it Quantum Optics}, 2nd ed. (Springer, Berlin, 2008).
\bibitem{Znidari} M. \v{Z}nidari\v{c}, T. Prosen, G. Benenti, G. Casati, and D. Rossini, Phys. Rev. E {\bf 81}, 051135 (2010).
\bibitem{Fubini} A. Fubini, G. Falci, and A. Osterloh, New J. Phys. {\bf 9}, 134 (2007); O. Viyuela, A. Rivas, and M. A. Martin-Delgado, Phys. Rev. B
{\bf 86}, 155140 (2012); A. Dhahri, J. Phys. A: Math. Theor. {\bf 41}, 275305 (2008).
\bibitem{Cardy} J. Cardy, {\it Scaling and Renormalization in Statistical Physics} (Cambridge University Press, Cambridge, U.K., 1996).
\bibitem{Sondhi} S. L. Sondhi, S. M. Girvin, J. P. Carini, and D. Shahar, Rev. Mod. Phys. {\bf 69}, 315 (1997); M. Vojta, Rep. Prog. Phys. {\bf 66}, 2069 (2003).
\bibitem{Coldea} R. Coldea, D. A. Tennant, E. M. Wheeler, E. Wawrzynska, D. Prabhakaran, M. Telling, K. Habicht, P. Smeibidl, and K. Kiefer, Science {\bf 327}, 177 (2010).


















\end{thebibliography}
\end{document}